\documentclass[conference]{IEEEtran}
\IEEEoverridecommandlockouts

\usepackage{cite}
\usepackage{amsmath,amssymb,amsfonts}
\usepackage{algorithmic}
\usepackage{graphicx}
\usepackage[dvipsnames]{xcolor}
\usepackage[final]{microtype}
\usepackage[italic]{mathastext}
\usepackage{libertine}
\usepackage[T1]{fontenc}
\usepackage{textcomp}
\usepackage[varqu,varl]{zi4}
\usepackage[all]{nowidow}
\usepackage{fancyhdr}
\usepackage{booktabs,multirow,tabularx}
\usepackage{enumitem}
\usepackage{comment}
\usepackage{hyperref}
\usepackage{fancybox}
\usepackage[breakable]{tcolorbox}

\def\BibTeX{{\rm B\kern-.05em{\sc i\kern-.025em b}\kern-.08em
    T\kern-.1667em\lower.7ex\hbox{E}\kern-.125emX}}

\begin{document}

\title{Agentic AI Workload Characterization}

\author{
\IEEEauthorblockN{
Yichao Yuan\IEEEauthorrefmark{1},
Ankita Nayak\IEEEauthorrefmark{2},
Souvik Kundu\IEEEauthorrefmark{3},
Nishil Talati\IEEEauthorrefmark{1}
}

\IEEEauthorblockA{
\IEEEauthorrefmark{1}
\textit{University of Illinois Urbana-Champaign, USA} \\
\{yichaoy2,nishil\}@illinois.edu
}

\IEEEauthorblockA{
\IEEEauthorrefmark{2}
\textit{Gimlet Labs}, USA \qquad
\IEEEauthorrefmark{3}
\textit{Intel Corporation}, USA\\
ankitanayak@gimletlabs.ai \qquad
souvikk.kundu@intel.com
}
}

\maketitle

\begin{abstract}
Agentic AI shifts LLM serving from isolated prompt-generation requests to stateful, multi-turn executions that repeatedly invoke the model, call tools, and grow context over time. This paper characterizes ReAct-style agents from both the LLM-serving and tool-execution perspectives using an end-to-end tracing infrastructure across reasoning and non-reasoning Gemma and Qwen configurations on five agentic benchmarks. Our study shows that agentic workloads are not simply long-prompt workloads: with effective context caching, most input tokens are reused across turns, making execution decode-dominated while increasing dependence on long-lived KV-cache state. We also find that tool use has a clear temporal structure, with agents shifting from read/explore behavior early in execution to execute/write behavior later. These results show that efficient agentic serving must jointly manage repeated model re-entry, persistent context state, and workload-dependent tool behavior.
\end{abstract}

\begin{IEEEkeywords}
Agentic AI, large language models, workload characterization, LLM serving, KV cache, tool use
\end{IEEEkeywords}

\section{Introduction}

Large Language Models (LLMs) have accelerated the broader impact of AI, transforming several domains, including scientific discovery and software engineering, from AlphaFold-style protein-structure prediction~\cite{jumper2021alphafold} to AI-assisted programming with tools such as GitHub Copilot~\cite{peng2023copilot}. The next frontier is \textbf{Agentic AI}~\cite{yao2023react,liu2023agentbench,ma2024agentboard}: systems that combine LLMs with tools, execution environments, memory, and control logic to pursue goals over multiple steps. Unlike a single LLM call, an agent repeatedly reasons, acts, observes outcomes, and adapts~\cite{yao2023react}, making it attractive for software engineering~\cite{deng2025swebenchpro}, data analytics~\cite{stancil2026adebench}, general-assistant workflows~\cite{mialon2023gaia}, and web-based tasks~\cite{yao2022webshop}.

Agentic systems span ReAct-style reason-act-observe loops~\cite{yao2023react}, Reflexion-style self-reflection~\cite{shinn2023reflexion}, Toolformer-style tool learning~\cite{schick2023toolformer}, and LATS-style planning/search~\cite{zhou2024lats}, differing in how they organize reasoning, tool use, observations, planning, and delegation. We focus on \textbf{ReAct-style agents} because they are simple, widely used, and closely match the serving loop of systems such as Claude Code~\cite{anthropic2026claudecode}, Codex~\cite{openai2026codex_agent_loop}, and OpenClaw~\cite{openclaw2026}. Prior work studies agent capability, progress, behavior, cost, and CPU-centric execution~\cite{liu2023agentbench,ma2024agentboard,kim2025costdynamicreasoning,raj2025cpucentricagentic}. However, they leave open a systems-level question: \textit{how do model serving, context growth, tool latency, and failure-driven retries interact to determine the efficiency bottlenecks of end-to-end agent execution?}

To address this gap, this paper characterizes ReAct-style agent workloads from both the LLM-serving and tool-execution perspectives. In specific, we develop a tracing infrastructure that runs Claude Code agents through Harbor~\cite{Harbor_Framework}, serves Gemma~\cite{google2026gemma4} and Qwen~\cite{qwen2026qwen36_27b} models with vLLM~\cite{kwon2023pagedattention}, and correlates agent trajectories with request-level serving traces. Our benchmark suite spans data analytics, general assistant tasks, software engineering, and terminal environments using ADE-Bench~\cite{stancil2026adebench}, DABstep~\cite{egg2025dabstep}, GAIA~\cite{mialon2023gaia}, SWE-bench Pro~\cite{deng2025swebenchpro}, and Terminal-Bench 2.0~\cite{merrill2026terminalbench}.

Our characterization shows that agentic workloads are long-tailed in two distinct ways: some tasks require many turns, while others accumulate large contexts with fewer turns. Surprisingly, \textit{reasoning does not always increase cost}; for example, for Gemma, thinking reduces pathological trajectories, with ADE averaging 18.0 turns for Gemma Thinking versus 108.8 turns and a 786-turn maximum for Gemma Instant. Meanwhile, \textit{context growth is not simply a proxy for turn count}: SWE-bench Pro has the largest accumulated contexts, averaging 69K--80K tokens, despite other workloads having more extreme turn-count tails. Thus, agentic serving must manage both repeated model re-entry and sustained KV-cache growth.

Our performance characterization shows that \textit{agentic LLM serving is not conventional long-prompt serving with an effective context cache design}. Most input tokens are reused across turns, yielding high empirical cache-hit ratios of 84.6--99.5\% and making decode dominate 91.0--98.6\% of LLM time. Thus, efficient agent serving depends less on optimizing isolated long prompts and more on preserving long-lived KV cache state across turns; if this state is evicted, a decode-dominated workload can become expensive recomputation.
While our analysis exposes the \textit{potential of high cache reuse} in agentic workloads, achieving such decode-dominated execution in practice can be challenging because concurrent agents with long multi-turn contexts can exceed GPU memory, requiring cache-retention, scheduling, offloading, or recomputation mechanisms that introduce additional overheads~\cite{li2025continuum,kang2026thunderagent,cheng2025lmcache,yuan2026kairos}.

Our tool characterization shows that tool execution is not a single workload category: tool calls range from lightweight file inspection to shell execution, web retrieval, delegated subagent work, and large returned observations. GAIA, for example, spends up to 28.7\% of time in tools due to WebFetch/WebSearch, while SWE-bench Pro and Terminal-Bench rely heavily on Bash, Read, Edit, Agent, and TaskOutput. \textit{Latency-heavy tools are not always failure-prone}: Agent and TaskOutput can dominate time and tokens, whereas Bash, Edit, and Read more often trigger retries, error observations, and context growth. 
In some cases, repeated tool failures can trap an agent in a recovery loop, causing long executions that consume turns and context without making meaningful progress.
We also analyze how tool interactions evolve over a task, finding that \textit{agents shift from early read/explore behavior for gathering information to later execute/write behavior for editing, testing, commands, and finalization.}

This work provides a unified, end-to-end characterization of agent behavior, context-state evolution, and serving-time performance using one tracing infrastructure built around Claude Code, a production-grade agent harness. Prior work studies only subsets of these dimensions and does not characterize them jointly. We also identify runtime-observable signals with direct performance implications, providing a practical basis for system optimization beyond behavioral observations. We summarize key differences in Table~\ref{tab:novelty_comparison}.
\begin{table}[t]
\centering
\caption{Comparison with closely related agent characterization studies.}
\label{tab:novelty_comparison}
\footnotesize
\setlength{\tabcolsep}{2pt}
\renewcommand{\arraystretch}{1.15}

\begin{tabularx}{\columnwidth}{
    @{}
    >{\raggedright\arraybackslash}X
    >{\centering\arraybackslash}p{0.11\columnwidth}
    >{\centering\arraybackslash}p{0.11\columnwidth}
    >{\centering\arraybackslash}p{0.11\columnwidth}
    >{\centering\arraybackslash}p{0.09\columnwidth}
    @{}
}
\toprule
\textbf{Category / Characterization} &
\textbf{Kim et al.~\cite{kim2025costdynamicreasoning}} &
\textbf{Raj et al.~\cite{raj2025cpucentricagentic}} &
\textbf{Ma et al.~\cite{ma2024agentboard}} &
\textbf{Ours} \\
\midrule

\textbf{Tool behavior}\\[-1pt]
{\scriptsize\itshape Latency, result size, failure behavior, and intention evolution}
&
&
&
&
\checkmark
\\

\midrule

\textbf{Agent execution behavior}\\[-1pt]
{\scriptsize\itshape Reasoning vs.\ non-reasoning model behavior}
&
&
&
&
\checkmark
\\

\midrule

\textbf{Agent execution behavior}\\[-1pt]
{\scriptsize\itshape Turn-count distribution and context growth across turns}
&
Both
&
Turn count only
&
Context growth only
&
Both
\\

\midrule

\textbf{Runtime performance}\\[-1pt]
{\scriptsize\itshape LLM vs.\ tool execution-time breakdown}
&
\checkmark
&
\checkmark
&
&
\checkmark
\\

\bottomrule
\end{tabularx}
\end{table}
In summary, this paper makes the following contributions:
\begin{itemize}
\item We develop an end-to-end tracing infrastructure to characterize agentic AI workloads across model serving and tool execution.

\item We characterize execution behavior, context growth, token reuse, and KV-cache dynamics across representative agent workloads.

\item We analyze the cost, latency, and failure behavior of tool interactions and their impact on end-to-end agent execution.
\end{itemize}

    
    

\section{Background} \label{section:background}

\begin{figure*}
    \centering
    \includegraphics[width=\linewidth,
    trim={0 3mm 0 0},
    clip]{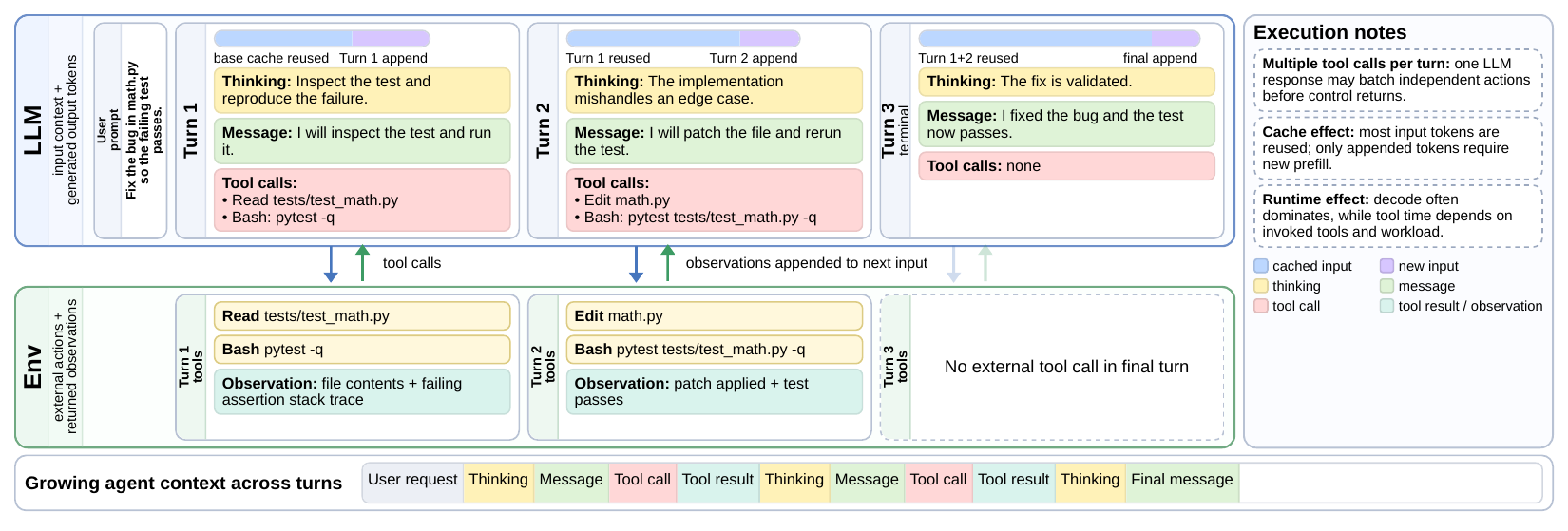}
    \caption{ReAct agent execution and context growth example. Each turn consists of an LLM response, one or more tool calls, and environment observations. Across turns, prior reasoning, messages, tool calls, and observations are appended to the agent context. If a tool call launches a subagent, we treat the subagent execution as part of the parent agent’s tool call.}    \label{fig:react-example}
\end{figure*}

\subsection{LLM Agents and the ReAct Paradigm}
LLM agents are systems that use LLMs to iteratively reason, plan, and interact with external tools or environments to accomplish user-specified goals~\cite{yao2023react,shinn2023reflexion,zhou2024lats,schick2023toolformer}. 
They shift the LLM serving paradigm from isolated single-turn prompt--generation requests to a closed-loop, multi-turn execution process, in which agents repeatedly interact with the model, external tools, and the environment over an evolving context~\cite{lin2024parrot,santhanam2024alto,asgar2025agentic,wadlom2026helium}.
Agents can follow different organizational patterns, which vary in how they orchestrate LLM invocations, incorporate environmental observations, and determine subsequent actions; representative examples include ReAct~\cite{yao2023react}, Reflexion~\cite{shinn2023reflexion}, Toolformer-style tool use~\cite{schick2023toolformer}, and LATS~\cite{zhou2024lats}.
Among these, ReAct is one of the most widely adopted paradigms due to its effectiveness and simplicity~\cite{yao2023react}. 
ReAct executes agents as a reason-act-observe loop: the LLM reasons about the next step, invokes tools or the environment, and incorporates the observation into later turns~\cite{yao2023react}. 
This makes agent execution a multi-turn process over an evolving context~\cite{li2025continuum,wadlom2026helium}.
It underlies many contemporary CLI agents, including Claude Code~\cite{anthropic2026claudecode} and Codex~\cite{openai2026codex_agent_loop}, and is the focus of this paper.

\subsection{Reasoning LLMs and Agents}

Reasoning LLMs are models trained to generate additional intermediate tokens for deliberation before producing a final response, which can improve their ability to solve complex tasks and increase task success rates~\cite{yao2023react,shinn2023reflexion,zhou2024lats}. 
This capability is now supported across a broad range of state-of-the-art models, from large proprietary systems such as GPT-5.5~\cite{openai2026gpt55} to small- and mid-sized open-weight models such as Gemma 4~\cite{google2026gemma4} and Qwen3.6~\cite{qwen2026qwen36_27b}.
Motivated by these gains, modern LLM agents increasingly incorporate reasoning LLMs to improve their effectiveness on complex, multi-step tasks~\cite{shinn2023reflexion,zhou2024lats,yang2024sweagent}. 
When used in ReAct agents, reasoning LLMs generate such intermediate tokens at every agent step before issuing tool calls or taking other actions~\cite{yao2023react}. 
These tokens are retained in the evolving context until the current user request is completed, thereby increasing both generation length and the agent's context usage, i.e., KV cache of conversation history~\cite{kwon2023pagedattention,cheng2025lmcache,li2025continuum}.

Figure~\ref{fig:react-example} illustrates a ReAct-style agent execution and the corresponding evolution of the serving context. 
At each agent step, the LLM consumes the accumulated context, generates intermediate reasoning tokens, a message, and optionally one or more tool calls; the environment then executes these calls and appends the returned observations to the next LLM input~\cite{yao2023react,schick2023toolformer}. 
A single LLM response may issue multiple tool calls before control returns to the model, causing multiple tool-use records and observations to be added within one step~\cite{schick2023toolformer}. 
As the agent progresses, most previous input tokens can be reused through the prefix/KV cache~\cite{vllm_apc}, while newly generated reasoning tokens, messages, tool calls, and tool results extend the context and increase the amount of state that must be maintained for subsequent steps~\cite{kwon2023pagedattention,cheng2025lmcache,li2025continuum}.

\subsection{Agentic Serving Terminologies}

Formally, we study the serving of ReAct-style agents powered by both reasoning and non-reasoning modes of LLMs.
We describe a ReAct-style agent as a multi-step interaction process~\cite{yao2023react}. 
For an agent $a$, let $H_{a,i}$ denote its accumulated context (trajectory) before the $i$-th LLM invocation, and let
\begin{equation}
C_{a,i} = |H_{a,i}|
\end{equation}
be the corresponding context length.

At step $i$, conditioned on $H_{a,i}$, the agent uses the \textit{LLM serving system} to generate an assistant-side response with a model that may contain three components:
\begin{equation}
z_{a,i} = (\theta_{a,i},\, m_{a,i},\, u_{a,i}),
\end{equation}
where $\theta_{a,i}$ denotes intermediate reasoning, or \emph{thinking}, tokens, $m_{a,i}$ denotes \textit{message} tokens, and $u_{a,i}$ denotes \textit{tool-call} tokens. 
Some components may be empty, and their exact ordering is model- and chat-template-dependent.
For non-reasoning models (no thinking),  $\theta_{a,i}=\varnothing$.

Before the next LLM invocation, these generated components are incorporated into the agent history according to the model's chat template. 
If tool calls are issued, the \textit{environment} executes them and appends the tool-use results $o_{a,i}$ to the history. 
The context therefore evolves as
\begin{equation}
H_{a,i+1}
=
H_{a,i}
\Vert
\Phi(\theta_{a,i},\, m_{a,i},\, u_{a,i})
\Vert
o_{a,i},
\end{equation}
where $\Phi(\cdot)$ denotes chat-template formatting and $\Vert$ denotes concatenation.
For a terminal step that completes normally, no tool calls or tool-use results are produced, i.e., $u_{a,i}=o_{a,i}=\varnothing$. 
We define an \emph{agent turn} as one interaction in which the agent issues tool call(s) and receives their results.

Following a common pattern in contemporary CLI agents such as Claude Code~\cite{anthropic2026claudecode_subagents}, an agent may invoke tools that launch additional agents.
We refer to the initiating agent as the \emph{main agent} and each launched agent as a \emph{subagent}. 
Consistent with Claude Code's convention, we do not allow nested subagent invocations in this work~\cite{anthropic2026claudecode_subagents}.

\section{Characterization Methodology} \label{section:methodology}

This work aims to comprehensively characterize the performance characteristics and behavioral patterns of real-world LLM agents across different tasks. 
The design space for agentic workloads is broad, spanning many agent frameworks, models, benchmarks, tools, and task domains, and an exhaustive study of this space is beyond the scope of a single paper. 
We therefore focus on a representative slice of this space, as detailed below. 
To this end, we evaluate agents under diverse model configurations, including reasoning and non-reasoning modes, and across multiple agentic benchmarks. 
We use Claude Code~\cite{anthropic2026claudecode} as the agent framework throughout our study. 
Concretely, our characterization spans the following dimensions:

\begin{itemize}[leftmargin=*]
    \item \textbf{Models.} We evaluate two open-weight LLMs: Qwen3.6-27B~\cite{qwen2026qwen36_27b} and Gemma4-31B~\cite{google2026gemma4}.

    \item \textbf{Reasoning configurations.} For each model, we evaluate both thinking and non-thinking modes.
    We denote the non-thinking mode as the model's \textit{Instant} variant, while using the raw model name for its thinking mode.
    The two modes are enabled by setting the corresponding knob in the model's chat template.

    \item \textbf{Benchmark datasets.} We evaluate agents across a diverse set of agentic workloads, including:
    \begin{itemize}[leftmargin=*]
        \item \textbf{ADE-Bench}~\cite{stancil2026adebench}: data analytics workloads,
        \item \textbf{DABStep*}~\cite{egg2025dabstep}, data agent tasks,
        \item \textbf{GAIA}~\cite{mialon2023gaia}, representing general-purpose assistant tasks that require multi-step reasoning and tool use,
        \item \textbf{SWE-bench Pro*}~\cite{deng2025swebenchpro}, software engineering tasks,
        \item \textbf{Terminal-Bench 2.0}~\cite{merrill2026terminalbench}: terminal-based software engineering and system interaction tasks.
    \end{itemize}
\end{itemize}
*For DABStep and SWE-bench Pro, the full benchmark suites are large and expensive to run end-to-end with multiple models and reasoning configurations. We therefore randomly sample 100 tasks from each benchmark suite to keep the study computationally tractable while still preserving task diversity within each workload. This allows us to collect fine-grained per-turn, per-request, and per-tool traces across all configurations, rather than limiting the characterization to a smaller set of models or metrics.

\subsection{Characterization Infrastructure}

We build an end-to-end characterization infrastructure to execute agents, serve models, and collect fine-grained request-level traces. 
All experiments are conducted on a server equipped with two NVIDIA H100 NVL GPUs connected by 12 NVLink links, and an Intel Xeon Platinum 8592+ CPU. 
We serve all models using vLLM v0.20.0~\cite{kwon2023pagedattention} from the official Docker image, with tensor parallelism set to 2. 
We enable vLLM's OpenTelemetry support~\cite{opentelemetry2026docs} and collect trace data through Jaeger~\cite{jaeger2026docs}, which provides detailed timing information for each request, including prefill and decode execution time.

We use Harbor's evaluation environment~\cite{Harbor_Framework}, which setup a single agent execution's objective and execution environment, to launch and evaluate agent tasks.
We use Claude Code~\cite{anthropic2026claudecode} as the agent scaffold.
To associate low-level LLM serving traces with high-level agent execution, we implement a customized request gateway between the agent and the vLLM server. 
The gateway has two components.

\begin{itemize}[leftmargin=*]
    \item \textbf{Forwarding proxy.} 
    The proxy forwards every LLM request issued by the agent to the vLLM server while capturing and logging the request metadata. 
    It also modifies each request to ensure that vLLM emits OpenTelemetry traces~\cite{opentelemetry2026docs} for that request, allowing us to correlate agent-level interactions with serving-level metrics such as prefill time, decode time, and request latency.

    \item \textbf{Per-agent wrapper.} 
    The wrapper launches each agent process and redirects its LLM traffic to the gateway. 
    For each agent execution process, the wrapper assigns a unique API key, which allows the gateway to group all requests belonging to the same agent execution. 
    This enables per-agent reconstruction of request sequences and supports subsequent analysis of both system performance and agent behavior.
\end{itemize}

\begin{figure}
    \centering
    \includegraphics[width=\linewidth]{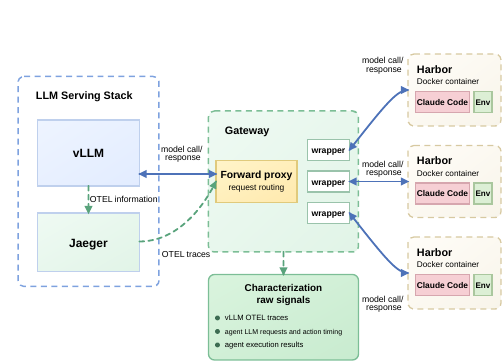}
    \caption{Tracing infrastructure for characterizing agent workload components and their interactions. 
Harbor launches Dockerized Claude Code agents, whose model traffic is redirected by a wrapper through our gateway. 
The gateway forwards requests to vLLM, records agent-side request and action timing, and attaches metadata for per-agent trace correlation. 
vLLM emits OpenTelemetry (OTEL) traces collected by Jaeger; these traces, together with gateway logs and Harbor execution results, form the raw signals used in our characterization.
    }
    \label{fig:infra}
\end{figure}

Figure~\ref{fig:infra} summarizes our characterization infrastructure. 
Harbor launches isolated Claude Code agent executions, whose LLM requests are redirected through a gateway that records agent-level metadata and forwards requests to vLLM. 
vLLM emits OpenTelemetry traces collected by Jaeger, and the characterization pipeline combines these serving traces with agent request logs, action timing, and execution results.

\section{Agent Execution Characterization} \label{section:execution_level_analysis}

This section characterizes how agentic workloads evolve over the task lifetime. We ask four related questions: (1) how many turns agents execute, (2) how much context they accumulate across those turns, (3) what is the composition of generated output tokens, and (4) whether successful and failed runs impose different context demands.

\begin{figure*}[t]
    \centering
    \includegraphics[width=\linewidth]{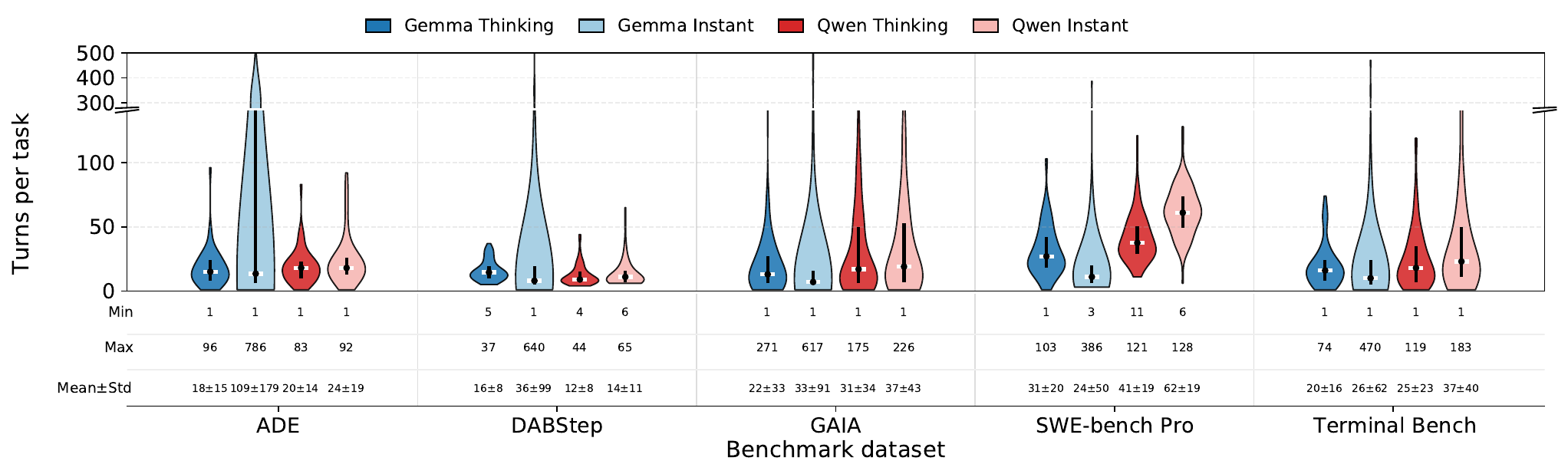}
    \caption{Distribution of agent turns per task. Agent trajectories are highly variable: instant variants often require more turns and exhibit longer tails, while reasoning variants generally produce more compact interactions.}
\label{fig:turn_count_distribution}
\end{figure*}
\noindent
\textbf{\underline{Agent Turn Counts.}}
Figure~\ref{fig:turn_count_distribution} shows that agent turn count is highly dependent on both workload and model behavior. Although most configurations average between 12--62 turns per task, several exhibit extreme long tails. Gemma Instant is the clearest example, reaching 786 turns on ADE, 640 on DABStep, 617 on GAIA, and 470 on Terminal Bench. In contrast, Gemma Thinking remains much more compact on most datasets, suggesting that explicit reasoning can reduce inefficient multi-turn exploration.

The benefit of reasoning is strongest for Gemma. On ADE, Gemma Thinking averages only 18.0±15.3 turns, while Gemma Instant averages 108.8±178.7 turns. Qwen shows a milder but still visible gap, with Qwen Instant usually requiring more turns than Qwen Thinking. SWE-bench Pro is a notable case: Qwen Instant averages 62.4 turns and shows a concentrated high-turn distribution, indicating consistently long trajectories rather than a few outliers. These results show that agentic execution lifetimes are both long-tailed and model-dependent, which has direct implications for serving systems because each additional turn requires re-entering the model, extending context, maintaining KV cache state, and scheduling another prefill/decode phase.

The extreme tail for ADE is driven primarily by the Gemma Instant configuration, where Table~\ref{tab:failing_tools} shows an unusually high number of failing `Edit' calls: 2,757 edit attempts with a 95.4\% failure rate. This suggests that \textit{the agent is not merely taking many productive steps, but is trapped in a repeated edit-failure pattern in which it attempts to modify files, receives failure/error observations, and then retries without successfully recovering.} This behavior is consistent with Figure~\ref{fig:turn_count_distribution}, where Gemma Instant on ADE reaches 786 turns and averages 108.8 turns, far above the corresponding Gemma Thinking configuration. Importantly, this pathological failure loop is not observed at the same severity in the other model--dataset combinations, indicating that the long tail is caused by a specific interaction between the non-thinking Gemma model and ADE’s edit-heavy task structure. Thus, \textit{the high turn count in this case reflects failure-driven retry behavior} rather than inherently longer successful task execution.

\begin{figure*}[ht]
    \centering
    \includegraphics[width=\linewidth]{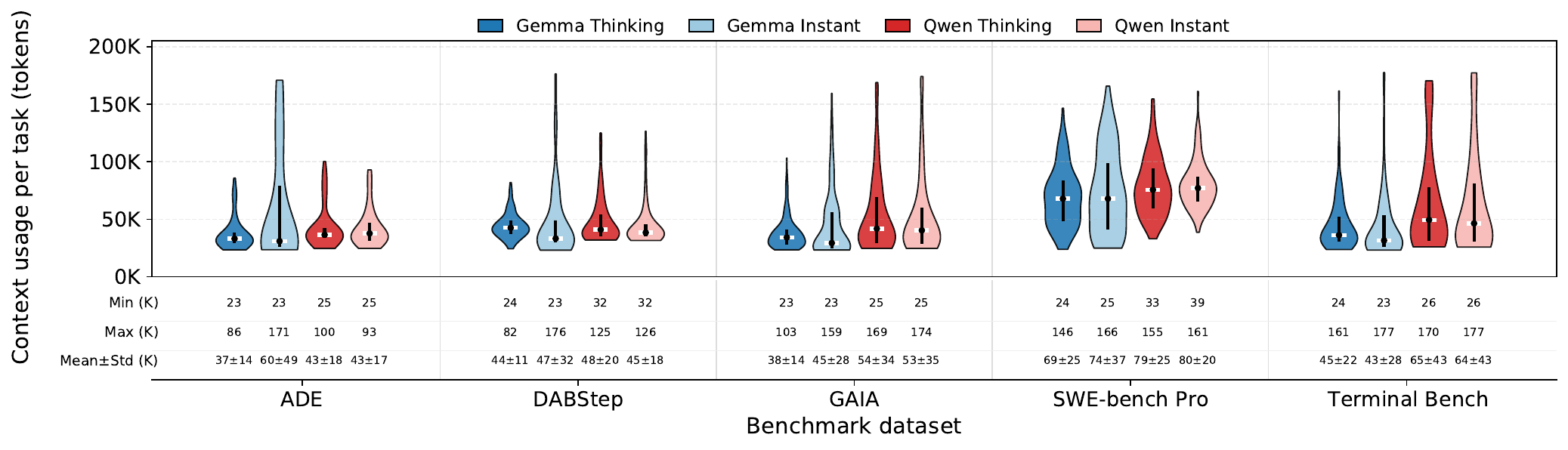}
    \caption{Distribution of accumulated context usage across all turns for each task. Context usage varies substantially by workload, with SWE-bench Pro consistently producing the largest contexts, while instant variants often introduce longer tails even when their medians remain close to reasoning variants.}
\label{fig:context_usage_violin}
\end{figure*}
\noindent
\textbf{\underline{Agent Context Length.}}
Figure~\ref{fig:context_usage_violin} shows the accumulated context usage across all turns of a task, measured in thousands of tokens. Unlike turn count in Figure~\ref{fig:turn_count_distribution}, context usage is less dominated by a few extreme trajectories and more consistently reflects workload complexity. \textit{SWE-bench Pro is the most context-heavy benchmark} across all configurations, with mean context usage between 68.7K and 80.1K tokens and maxima between 146K and 166K tokens. Terminal Bench and GAIA also produce large contexts for Qwen, reaching mean usages of 63.5K--65.1K on Terminal Bench and 52.5K--54.5K on GAIA, while ADE and DABStep are smaller except for long-tail Gemma Instant cases.

The comparison between reasoning and instant variants shows that instant models do not always increase average context usage, but they often increase variance and tail behavior. Gemma Instant is especially variable: on ADE, it averages 60.3K tokens with a maximum of 171K, compared to 37.2K and 86K for Gemma Thinking; similar long tails appear on DABStep and GAIA. Qwen behaves differently: the thinking and instant variants have similar context usage on several datasets. However, Qwen is consistently more context-heavy than Gemma on GAIA and Terminal Bench. 

Together with Figure~\ref{fig:turn_count_distribution}, this suggests that \textit{long agent trajectories and large accumulated contexts are related but not identical phenomena.} Some workloads require many model invocations because the agent proceeds through a long sequence of actions, while others accumulate large contexts because each turn carries substantial prompt, observation, tool-output, or intermediate reasoning state. This distinction is important because the two behaviors stress different parts of the serving stack: high turn counts increase scheduling frequency and repeated prefill/decode cycles, whereas large accumulated contexts increase KV-cache footprint, memory pressure, and context-reuse overheads. As a result, agentic serving systems must jointly optimize for repeated model re-entry and sustained context growth rather than treating sequence length or request count in isolation.

\begin{figure}[t]
    \centering
    \includegraphics[width=\linewidth]{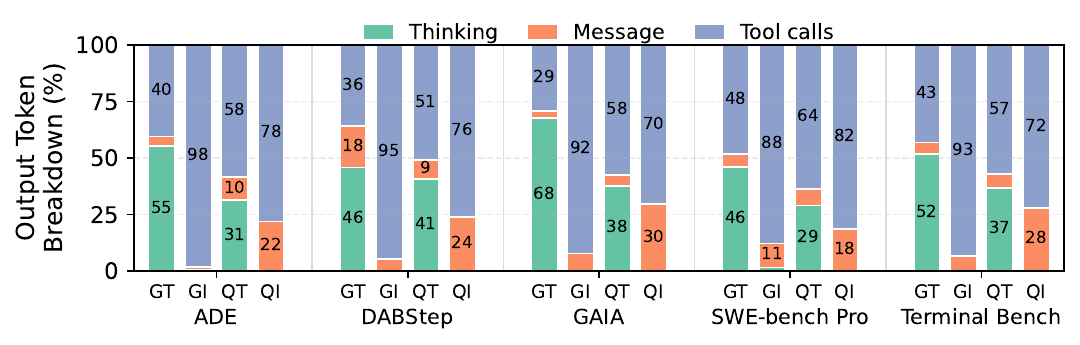}
    \caption{Breakdown of generated output tokens into thinking, message, and tool-call tokens. Reasoning variants spend a large fraction of their output budget on thinking, while instant variants are dominated by tool-call generation.}
\label{fig:output_token_breakdown_horizontal}
\end{figure}
\noindent
\textbf{\underline{Agent Output Token Composition.}}
Figure~\ref{fig:output_token_breakdown_horizontal} shows how each model’s generated output tokens are distributed across thinking tokens, normal message tokens, and tool-call tokens. For reasoning models, thinking tokens form a substantial fraction of generation: Gemma Thinking spends 45.8--67.6\% of its output on thinking across workloads, while Qwen Thinking spends 29.0--40.7\%. This confirms that enabling reasoning changes not only total token volume, but also the semantic composition of generated tokens. In contrast, instant variants generate no thinking tokens, so their output is split almost entirely between user-visible messages and tool calls.

An interesting result is that, the \textit{tool-call tokens dominate output generation in many settings, especially for instant variants}. Gemma Instant spends 87.8--98.2\% of its output tokens on tool calls, while Qwen Instant spends 70.4--81.6\%. Even for reasoning variants, tool calls remain a large component: Qwen Thinking spends 50.8--63.7\% of its output on tool calls in four of the five workloads, and Gemma Thinking reaches 48.2\% on SWE-bench Pro. This shows that \textit{ReAct-style agents are not simply producing long natural-language responses; much of their generated output is structured action generation that drives interaction with the environment}. 

\begin{figure*}[t]
    \centering
    \includegraphics[width=\linewidth]{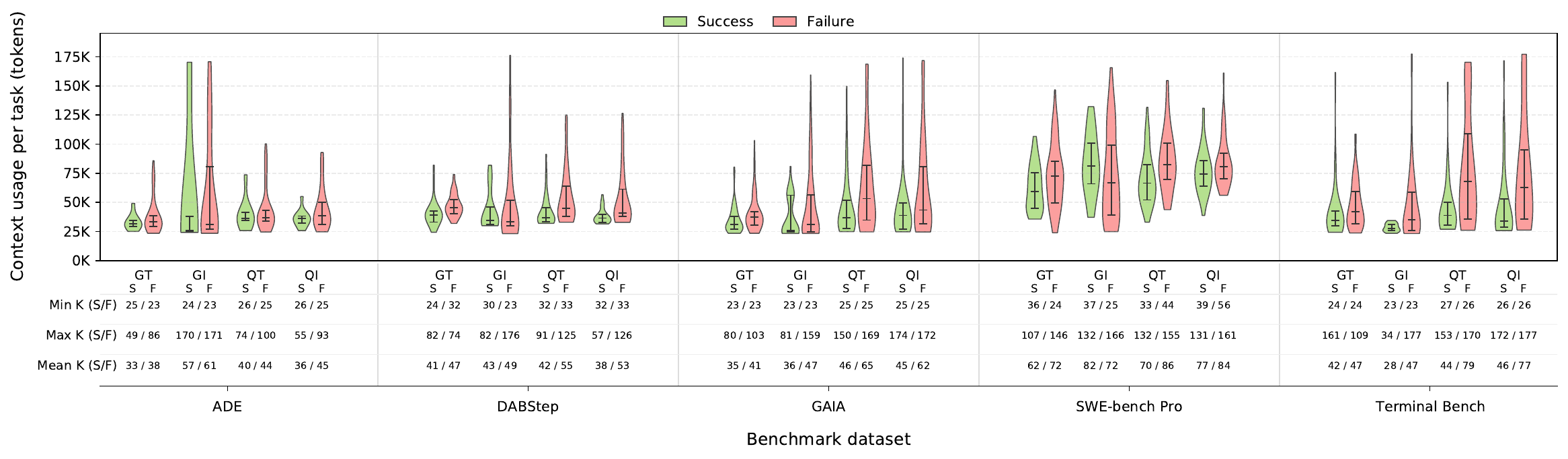}
    \caption{Context-usage distributions split by task outcome. Failed agents often carry larger contexts (up to 1.8$\times$ in terms of mean) and longer tails, that runs can increase memory pressure even when they do not produce useful task outcomes.}
\label{fig:success_fail_context_usage_violin}
\end{figure*}
\noindent
\textbf{\underline{Successful Versus Failed Agent Context Lengths.}}
Figure~\ref{fig:success_fail_context_usage_violin} compares accumulated context usage for successful and failed tasks. The overall trend is that failed agents usually accumulate larger contexts than successful agents, although the strength of this effect varies by workload and model. On GAIA, failures increase mean context usage for every configuration: from 34.6K to 40.6K tokens for Gemma Thinking, 36.4K to 47.3K for Gemma Instant, 45.6K to 64.7K for Qwen Thinking, and 44.9K to 62.4K for Qwen Instant. Terminal Bench shows an even sharper gap for Qwen, where failed runs reach 79.2K and 76.6K tokens on average for Qwen Thinking and Qwen Instant, compared to 44.4K and 45.9K for successful runs.

Failed agents often take more turns and also carry more accumulated context. The mechanism is consistent with Table~\ref{tab:failing_tools}: failed Bash, Edit, or Read actions append error messages, stack traces, failed patches, or diagnostic outputs into the agent history, which then become part of the next turn's input context. As a result, failed runs can amplify system load in two dimensions at once: they invoke the model more times and they increase the KV cache footprint carried across those invocations.

There are also useful exceptions. SWE-bench Pro is context-heavy for both successful and failed agents, and Gemma Instant has a higher mean context for successful runs than failed runs, 82.2K versus 72.3K tokens. This suggests that \textit{large context is not inherently a failure mode}; some successful software-engineering tasks genuinely require sustained context about code, tests, and edits. The broader implication is that \textit{outcome alone does not determine serving cost, but failed agents are a major source of long-tail context growth and therefore should be included explicitly when characterizing agentic serving workloads.}

\textit{\noindent\textbf{\underline{Summary of Insights.}}
Agentic workloads are long-tailed in two distinct ways: some runs execute many turns, while others accumulate large contexts even with fewer turns. Surprisingly, reasoning does not uniformly increase workload cost; for Gemma, reasoning substantially reduces pathological long-turn trajectories, while instant variants often create longer tails. Failures also amplify system load: unsuccessful agents often execute more turns and carry larger contexts, meaning failed tasks can be expensive even when they do not produce useful outcomes.}
\section{Runtime Performance Characterization} \label{section:behavior_analysis}

This section characterizes where agent execution time is spent and how context reuse shapes LLM serving cost. We ask four related questions: (1) how much time agents spend in LLM inference versus external tool execution, (2) how much of each turn's input context is reused through prefix/context caching versus newly appended, (3) whether LLM time is dominated by prefill or decode, and (4) how context-cache thrashing under memory pressure affects cache effectiveness and serving throughput.

\begin{figure}[t]
    \centering
    \includegraphics[width=\linewidth]{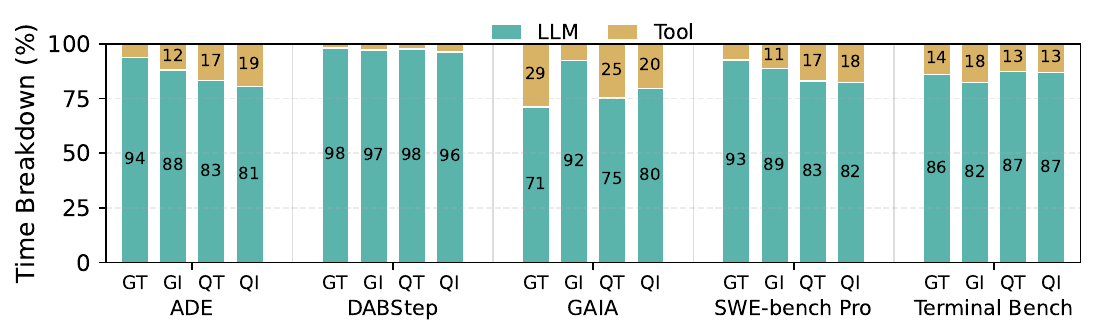}
    \caption{Breakdown of end-to-end execution time between LLM inference and tool execution. Across workloads, most time is spent inside the LLM, but tool time becomes non-negligible for GAIA and Terminal Bench, reflecting stronger dependence on external interaction.}
\label{fig:execution_component_breakdown}
\end{figure}
\noindent
\textbf{\underline{LLM Versus Tool Call Time.}}
Figure~\ref{fig:execution_component_breakdown} decomposes total agent execution time into LLM inference and tool execution. 
Across all benchmarks and models, execution is primarily LLM-dominated, with LLM inference accounting for 71--98\% of total runtime. 
DABStep is the most LLM-bound workload, spending 96--98\% of time in the model, while ADE shows a similar trend with a larger but still secondary tool component.

However, tool execution is not negligible. 
Tools account for 2--29\% of total runtime, with the largest fractions on GAIA, where tool time reaches 28.7\% for Gemma Thinking and 24.9\% for Qwen Thinking. 
This aligns with Table~\ref{tab:rq9_rq11_combined_top_tools}, where GAIA includes expensive WebFetch, WebSearch, Agent, and TaskOutput calls. 
SWE-bench Pro and Terminal Bench also incur non-trivial tool costs, reaching 17.7\% and 17.6\%, respectively. 
Overall, these results show that \textit{ReAct-style agent execution is dominated by repeated LLM invocations, but tool latency remains a substantial, workload-dependent component that serving systems should account for when scheduling, batching, or overlapping agent execution.}
While the degree of concurrency may change the LLM v.s. tool time, our results are in line with other works~\cite{luo2026cpugpucodesignagentic}.

\begin{figure*}[t]
    \centering
    \includegraphics[width=\linewidth]{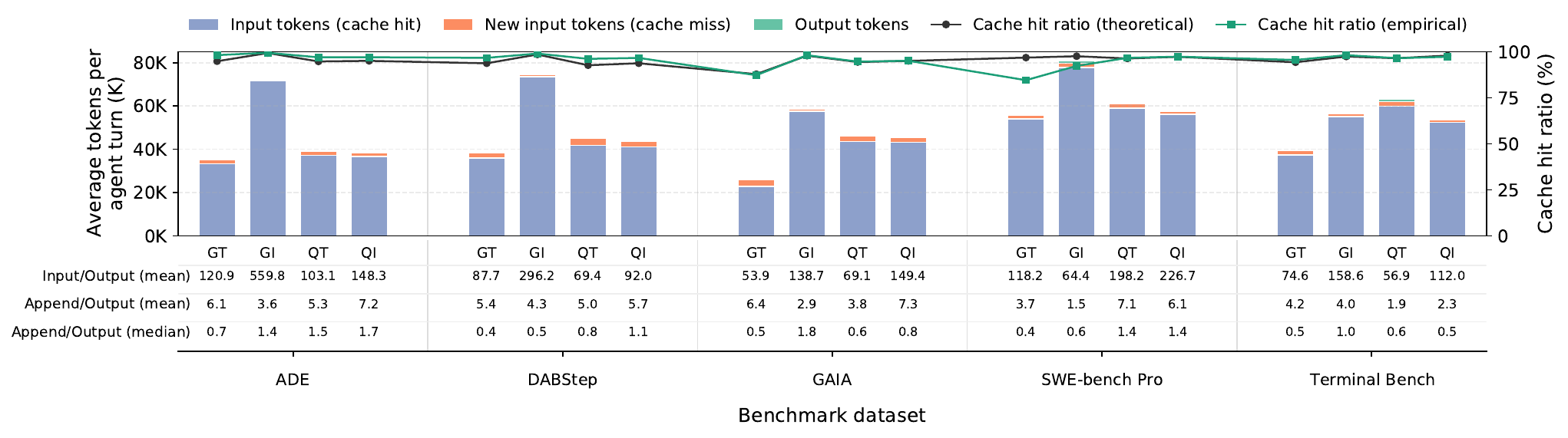}
    \caption{Breakdown of per-turn context into cached input tokens, newly appended input tokens, and output tokens. Most input tokens are served from the prefix/context cache, yielding high cache-hit ratios, but each turn still appends a small stream of new tokens that must be processed and incorporated into the growing agent state. Here we investigate workload's intrinsic characteristic independent of whether it thrashes (i.e., context running out of GPU memory) in reality.}
\label{fig:context_cache_breakdown}
\end{figure*}
\noindent
\textbf{\underline{Context Cache Effectiveness.}}
Figure~\ref{fig:context_cache_breakdown} decomposes the average tokens processed per agent turn into three components: input tokens that hit in the cache, newly appended input tokens that miss in the cache (i.e., cold miss), and generated output tokens. The dominant component is cached input context, which accounts for the vast majority of per-turn token volume in nearly every setting. The theoretical cache-hit ratio is consistently high, ranging from 87.9--99.3\%, and the empirical cache-hit ratio is similarly high, ranging from 84.6--99.5\%. This confirms that agentic execution contains substantial prefix reuse: each turn reuses most of the accumulated history while only appending a relatively small amount of new information. Note that the measured hit rate is close to our estimation, and slight deviation is due to factors like system prompt sharing, inconsistency in tool call parser and chat template, etc.

Raw input-to-output ratios can mischaracterize agent workloads as overwhelmingly prefill-heavy, because the raw input includes accumulated history that is largely reused across turns. 
While the mean input-to-output ratio ranges from 53.9$\times$ to 559.8$\times$, the mean append-to-output ratio is only 1.5$\times$--7.3$\times$, with medians often below 1.5$\times$. 
Thus, the relevant per-turn prefill pressure is better captured by incremental appends rather than the full input sequence. 
Conceptually, this means \textit{agent execution should not be modeled as a sequence of independent long-prompt requests, but as repeated model re-entry over a growing cached context.}
Note that the cache hit rates we measure are close to the ideal rates expected in the absence of context-cache thrashing. With sufficient GPU memory, active contexts remain resident and cached tokens can be reused across turns; under memory pressure. However, thrashing repeatedly evicts useful cached prefixes from GPU memory, substantially lowering the effective hit rate and increasing prefill overhead.

\begin{figure}[t]
    \centering
    \includegraphics[width=\linewidth]{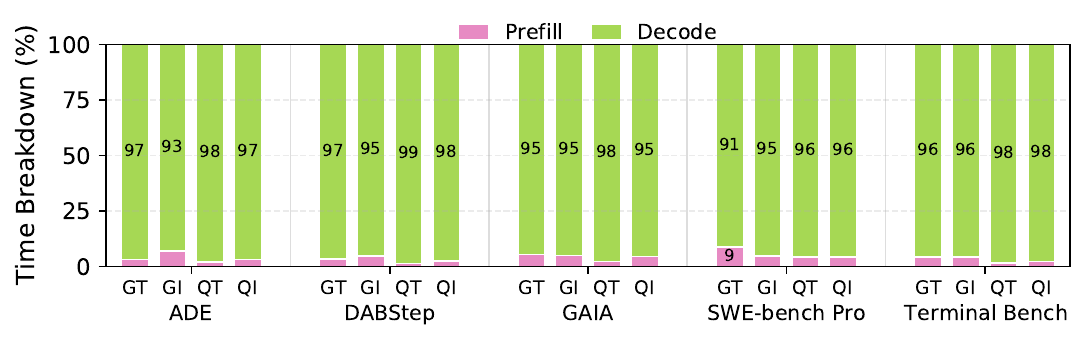}
     \caption{Breakdown of LLM execution time into prefill and decode phases. Despite large accumulated contexts, decode dominates runtime because prefix/context caching turns most of the input into reusable state and leaves only a small newly appended region to prefill at each turn. We control the concurrency s.t. no context thrashing happens.}
\label{fig:prefill_decode_breakdown_horizontal}
\end{figure}
\noindent
\textbf{\underline{Prefill Versus Decode Time.}}
Figure~\ref{fig:prefill_decode_breakdown_horizontal} decomposes LLM execution time into prefill and decode phases. 
Across all benchmarks and models, LLM execution is strongly decode-dominated: decode accounts for 91.0--98.6\% of LLM time, while prefill contributes only 1.4--9.0\%. 
This is consistent with the high context reuse shown in Figure~\ref{fig:context_cache_breakdown}: although agents accumulate long contexts over many turns, most input tokens are reused from the cache, so each turn only prefills the appended region rather than the full history.
We restate the same assumption here: this breakdown reflects a no-thrashing cache regime, where accumulated KV state over past conversations remains resident and is not evicted, recomputed, or fetched from slower memory tiers.

The largest prefill shares appear for Gemma Thinking on SWE-bench Pro and Gemma Instant on ADE, where prefill reaches 9.0\% and 7.1\%, respectively. 
Even in these cases, prefill remains a small fraction of LLM time because the append-to-output ratio is much lower than the raw input-to-output ratio: Figure~\ref{fig:context_cache_breakdown} shows mean input-to-output ratios up to 559.8$\times$, but mean append-to-output ratios of only 1.5$\times$--7.3$\times$. 
These results show that ReAct-style agent serving is better characterized as repeated decode over a growing cached context, rather than as repeated long-prompt prefilling.

\noindent
\textbf{\underline{Impact of Context Thrashing}}
We quantify context thrashing by increasing concurrent agent
jobs from Terminal-Bench and DABStep on a Qwen 3.6/H100
serving instance. The tables report prefix-cache hit rate,
serving throughput, and request time spent in LLM prefill.

\begin{table}[t]
    \centering
    \caption{Impact of context thrashing with increasing agent
    concurrency on Qwen3.6-27B serving. $\dagger$ denotes
    configurations where context thrashing occurs. Prefill time
    reports both absolute time and its fraction of request time.}
    \label{tab:context_thrashing}
    \footnotesize
    \setlength{\tabcolsep}{2.5pt}
    \renewcommand{\arraystretch}{1.05}
    \begin{tabularx}{\columnwidth}{@{}Xccc@{}}
        \toprule
        \multicolumn{4}{c}{\textbf{Terminal-Bench 2.0}} \\
        \midrule
        Max concurrency
            & 15 & 20$^\dagger$ & 25$^\dagger$ \\
        Prefix-cache hit (\%)
            & 98.85 & 16.78 & 19.21 \\
        Gen. throughput (tok/s/req.)
            & 30.08 & 16.96 & 10.30 \\
        Prefill time (s, \%)
            & 0.51 (4.42) & 2.24 (8.38) & 3.46 (7.81) \\
        \midrule
        \multicolumn{4}{c}{\textbf{DABStep}} \\
        \midrule
        Max concurrency
            & 5 & 10$^\dagger$ & 15$^\dagger$ \\
        Prefix-cache hit (\%)
            & 98.53 & 25.17 & 26.60 \\
        Gen. throughput (tok/s/req.)
            & 34.13 & 22.21 & 14.82 \\
        Prefill time (s, \%)
            & 0.36 (5.07) & 1.14 (6.57) & 3.10 (7.27) \\
        \bottomrule
    \end{tabularx}
\end{table}

Context thrashing first appears as a sharp prefix-cache hit-rate
drop, from about 98\% to 16--26\%, shifting the workload toward
prefill. It has two effects: (a) extra prefill tokens slow decode
for requests in generation, reducing average generation throughput
by up to 66\%; and (b) prefill time increases severalfold, though
it remains below 10\% of total request time. Thus, thrashing adds
overhead through longer prefill and, more importantly, slower
generation for inference batches. Our results assume recomputation
under thrashing. Retrieving evicted KV states from secondary tiers,
as in LMCache-style systems, is an alternative, but prior
work shows such transfers can also be costly due to inter-tier
bandwidth limits and retrieval latency~\cite{cheng2025lmcache}.

\textit{\noindent\textbf{\underline{Summary of Insights.}}
Context caching can be a very effective technique to reduce the cost of prefill in agentic LLM serving. With ideal context caching, most input tokens are reused across turns, so the workload becomes dominated by the LLM decode phase rather than repeated full-context prefill. The implication is that efficient agent serving depends less on optimizing isolated long prompts and more on preserving long-lived KV/cache state across repeated model re-entry; losing that state would turn a decode-dominated workload into expensive recomputation.}
\section{Tool Call Characterization} \label{section:system_level_analysis}

This section characterizes how agents interact with external tools and how those interactions shape execution cost. We ask five related questions: (1) which tool types dominate agent behavior, (2) what concrete commands are issued through Bash, (3) which tools contribute the largest execution latency and returned observations, (4) which tools fail most often, and (5) how tool intent shifts over the course of an agent trajectory.

\begin{figure*}[t]
    \centering
    \includegraphics[width=\linewidth]{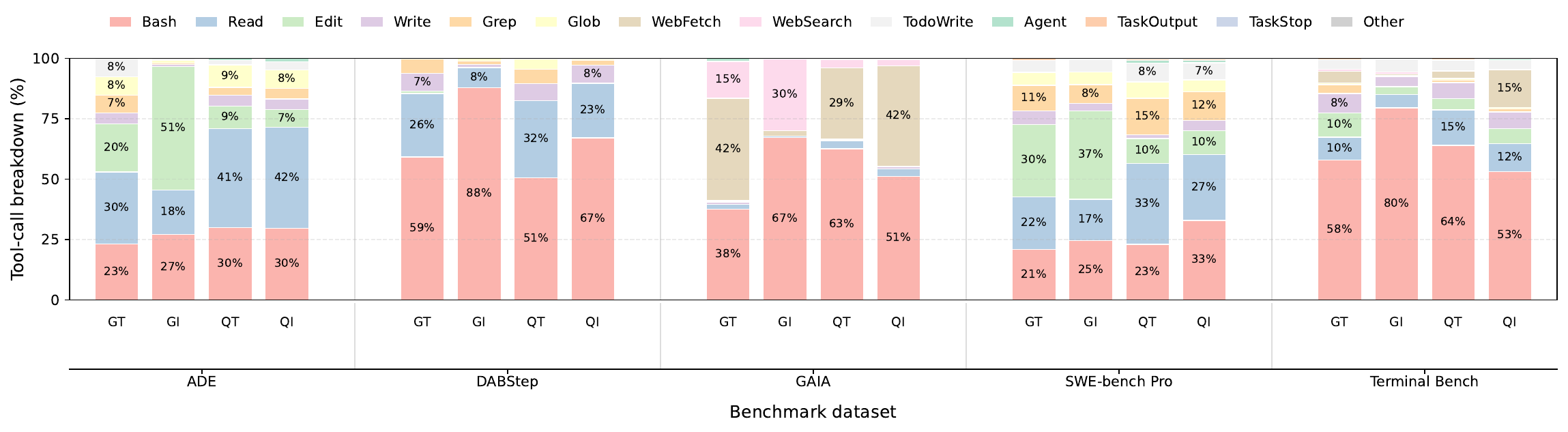}
    \caption{Breakdown of tool-call types across workloads. Tool usage is highly model and domain-dependent: coding and terminal tasks are dominated by Bash, Read, and Edit operations, while GAIA introduces substantial WebFetch/WebSearch activity, reflecting its stronger dependence on external information retrieval.}
\label{fig:tool_name_breakdown_stacked}
\end{figure*}
\noindent
\textbf{\underline{Breakdown by Tool Call Types.}}
Figure~\ref{fig:tool_name_breakdown_stacked} shows that tool-call composition varies much more by workload than by reasoning mode alone. ADE is dominated by repository-inspection and modification tools: Read accounts for 29.8--41.7\% of tool calls in three configurations, while Bash contributes 23.2--30.0\%. The main exception is Gemma Instant, where Edit alone accounts for 51.2\% of all tool calls. DABStep and Terminal Bench are more execution-heavy. Bash accounts for 50.5--87.8\% of tool calls on DABStep and 53.2--79.5\% on Terminal Bench, indicating that these tasks rely heavily on command-line execution rather than a broad mix of specialized tools.

GAIA has a qualitatively different profile because web-facing tools become major components of agent behavior. WebFetch accounts for 42.3\% of Gemma Thinking calls, 29.4\% of Qwen Thinking calls, and 41.6\% of Qwen Instant calls, while WebSearch reaches 29.7\% for Gemma Instant. This explains why GAIA had the largest tool-time fraction in Figure~\ref{fig:execution_component_breakdown}: its tools are not merely frequent, but also include retrieval-oriented operations that can be slower and more variable than local file or shell operations. In contrast, SWE-bench Pro is more balanced across code-editing actions, with Read, Edit, Bash, and Grep all contributing substantially; for example, Gemma Thinking spends 29.9\% of calls on Edit, 21.7\% on Read, and 21.0\% on Bash, while Qwen Instant spends 32.9\% on Bash, 27.3\% on Read, and 11.7\% on Grep.

The broader insight is that “tool use” is not a single systems category. A workload with many Bash calls stresses process execution and command-output handling, a workload with many Read/Edit/Grep calls stresses file-system interaction and codebase navigation, and a workload with WebFetch/WebSearch calls introduces external latency and retrieval variability. This distinction is important when interpreting earlier figures: similar tool-call counts per turn can correspond to very different runtime costs and context growth depending on which tools are invoked and how large their returned observations are.

\begin{figure*}[t]
    \centering
    \includegraphics[width=\linewidth]{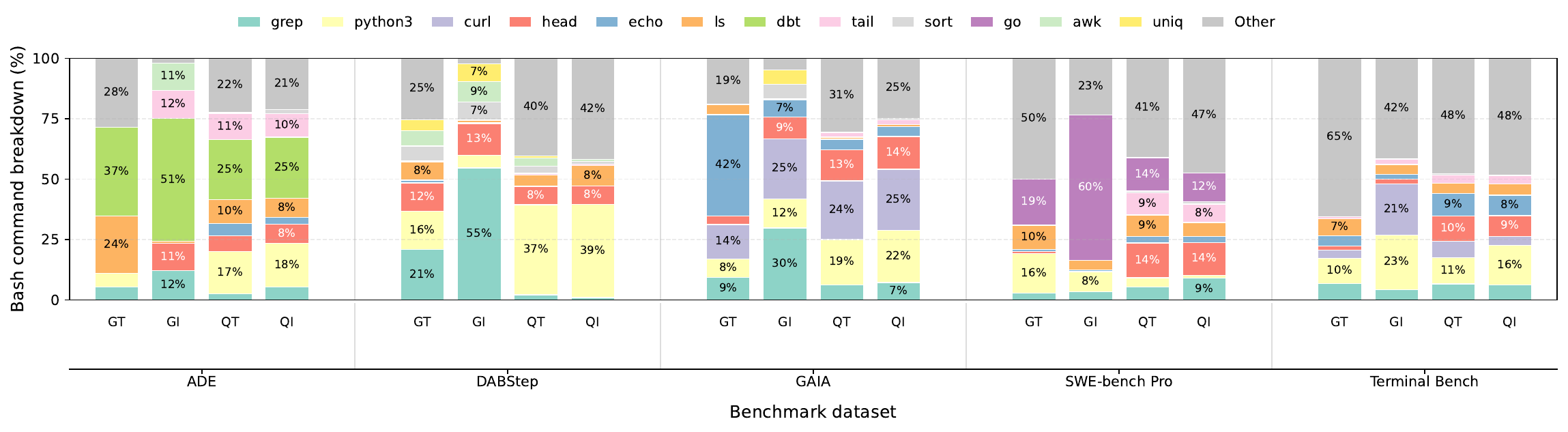}
    \caption{Breakdown of Bash commands issued by agents. Command usage reflects the task domain: database-oriented workloads rely heavily on dbt, web/retrieval workloads use curl and inspection commands, and coding/terminal workloads use a broader mix of language runtimes, testing commands, and shell utilities.}
\label{fig:bash_command_breakdown_stacked}
\end{figure*}
\noindent
\textbf{\underline{Breakdown of Bash Command Types.}}
Figure~\ref{fig:bash_command_breakdown_stacked} breaks down Bash invocations by the concrete command executed. The key observation is that Bash is not a monolithic tool: the command mix changes substantially across workloads, revealing what the agent is actually doing after it decides to execute code or interact with the shell. ADE is dominated by dbt, which accounts for 24.7--50.9\% of Bash commands across configurations, consistent with database-transformation style tasks. DABStep has a different structure: Gemma Instant uses grep for 54.7\% of Bash commands, whereas Qwen relies heavily on python3, which accounts for 37.2--38.5\% of its Bash commands. This suggests that even when agents operate on the same benchmark, different models may choose different execution strategies: one may inspect text through shell utilities, while another may script more of the task logic.

GAIA and the coding/terminal benchmarks expose another dimension of command diversity. GAIA includes a substantial share of curl commands, reaching 24.2--25.2\% for Qwen and 24.8\% for Gemma Instant, aligning with the earlier tool-type breakdown where web-facing tools were prominent. SWE-bench Pro and Terminal Bench have large Other fractions, reaching 50.0\% for Gemma Thinking on SWE-bench Pro and 65.5\% for Gemma Thinking on Terminal Bench. This category is not a single command; it aggregates locally rare commands, parsing artifacts such as `parse\_error' and `empty', and global tail commands outside the top 12 shown in the legend. In practice, it includes meaningful shell, build, package, and workload-specific commands such as `python', `git', `find', `cat', `pip', `npm', `node', `make', `gcc', `g++', `ffmpeg', `pdflatex', and `duckdb', so a large Other share indicates command diversity rather than simply noise. 

The broader insight is that shell execution should be interpreted through the commands being run, not only through the number of Bash tool calls. A Bash call that runs grep, head, or cat is primarily an inspection operation, while one that runs python3, go, make, or benchmark-specific commands may execute tests, scripts, or workload logic and can produce very different latency and output-token footprints. This helps connect Figure~\ref{fig:bash_command_breakdown_stacked} back to Figure~\ref{fig:tool_name_breakdown_stacked}: even when Bash dominates the tool mix, the underlying systems behavior can vary from file inspection to program execution, package/build activity, or data retrieval.

\begin{table*}[t]
\centering
\scriptsize
\setlength{\tabcolsep}{2.7pt}
\renewcommand{\arraystretch}{1.10}
\caption{Top time-consuming tools. Each cell reports the tool name, call count $n$, mean duration $\mu$, and mean result length $r$, with linearly interpolated first and third quartiles shown as $[Q_1,Q_3]$. $\dagger$ marks tools that are also among the top-5 by result length.}
\label{tab:rq9_rq11_combined_top_tools}
\begin{tabularx}{\textwidth}{llXXXX}
\toprule
\textbf{Benchmark} & \textbf{Rank} & \textbf{Gemma-T} & \textbf{Gemma-I} & \textbf{Qwen-T} & \textbf{Qwen-I} \\
\midrule

\multirow{4}{*}{\textbf{ADE}}
& Top 1
& \shortstack[l]{\textbf{Bash}$^{\dagger}$\\{\scriptsize $n$=208}\\{\scriptsize $\mu$=2.43s [0.09, 2.67]}\\{\scriptsize $r$=802 [32, 616]}}
& \shortstack[l]{\textbf{Bash}$^{\dagger}$\\{\scriptsize $n$=1462}\\{\scriptsize $\mu$=2.56s [2.55, 2.71]}\\{\scriptsize $r$=480 [111, 513]}}
& \shortstack[l]{\textbf{Agent}$^{\dagger}$\\{\scriptsize $n$=16}\\{\scriptsize $\mu$=66.31s [44.28, 82.19]}\\{\scriptsize $r$=1779 [1259, 2438]}}
& \shortstack[l]{\textbf{Agent}$^{\dagger}$\\{\scriptsize $n$=35}\\{\scriptsize $\mu$=56.68s [32.31, 65.67]}\\{\scriptsize $r$=1626 [946, 2086]}} \\

& Top 2
& \shortstack[l]{\textbf{Grep}$^{\dagger}$\\{\scriptsize $n$=66}\\{\scriptsize $\mu$=0.08s [0.07, 0.09]}\\{\scriptsize $r$=333 [33, 276]}}
& \shortstack[l]{\textbf{Write}\\{\scriptsize $n$=41}\\{\scriptsize $\mu$=0.08s [0.05, 0.06]}\\{\scriptsize $r$=57 [44, 49]}}
& \shortstack[l]{\textbf{Bash}$^{\dagger}$\\{\scriptsize $n$=611}\\{\scriptsize $\mu$=1.27s [0.10, 2.64]}\\{\scriptsize $r$=402 [64, 434]}}
& \shortstack[l]{\textbf{Bash}$^{\dagger}$\\{\scriptsize $n$=747}\\{\scriptsize $\mu$=1.33s [0.11, 2.72]}\\{\scriptsize $r$=413 [58, 466]}} \\

& Top 3
& \shortstack[l]{\textbf{Write}\\{\scriptsize $n$=41}\\{\scriptsize $\mu$=0.07s [0.05, 0.09]}\\{\scriptsize $r$=48 [41, 46]}}
& \shortstack[l]{\textbf{Edit}$^{\dagger}$\\{\scriptsize $n$=2757}\\{\scriptsize $\mu$=0.07s [0.05, 0.08]}\\{\scriptsize $r$=123 [53, 129]}}
& \shortstack[l]{\textbf{TodoWrite}\\{\scriptsize $n$=39}\\{\scriptsize $\mu$=0.10s [0.04, 0.06]}\\{\scriptsize $r$=40 [37, 37]}}
& \shortstack[l]{\textbf{WebSearch}$^{\dagger}$\\{\scriptsize $n$=1}\\{\scriptsize $\mu$=0.15s [0.15, 0.15]}\\{\scriptsize $r$=233 [233, 233]}} \\

& Token-only extras
& {\scriptsize Read ($r$=718)}
& {\scriptsize Glob ($r$=102)}
& {\scriptsize Read ($r$=441)}
& {\scriptsize Read ($r$=459), Glob ($r$=290)} \\

\midrule

\multirow{4}{*}{\textbf{GAIA}}
& Top 1
& \shortstack[l]{\textbf{Agent}$^{\dagger}$\\{\scriptsize $n$=51}\\{\scriptsize $\mu$=321.57s [107.08, 521.74]}\\{\scriptsize $r$=253 [176, 304]}}
& \shortstack[l]{\textbf{Agent}$^{\dagger}$\\{\scriptsize $n$=6}\\{\scriptsize $\mu$=31.87s [24.83, 39.12]}\\{\scriptsize $r$=548 [291, 843]}}
& \shortstack[l]{\textbf{Agent}$^{\dagger}$\\{\scriptsize $n$=11}\\{\scriptsize $\mu$=916.04s [207.90, 1262.39]}\\{\scriptsize $r$=420 [0, 671]}}
& \shortstack[l]{\textbf{Agent}$^{\dagger}$\\{\scriptsize $n$=13}\\{\scriptsize $\mu$=650.73s [125.48, 929.74]}\\{\scriptsize $r$=650 [287, 781]}} \\

& Top 2
& \shortstack[l]{\textbf{WebFetch}$^{\dagger}$\\{\scriptsize $n$=1810}\\{\scriptsize $\mu$=29.76s [9.59, 33.84]}\\{\scriptsize $r$=67 [29, 64]}}
& \shortstack[l]{\textbf{WebFetch}$^{\dagger}$\\{\scriptsize $n$=123}\\{\scriptsize $\mu$=3.98s [0.79, 6.58]}\\{\scriptsize $r$=58 [33, 61]}}
& \shortstack[l]{\textbf{TaskOutput}$^{\dagger}$\\{\scriptsize $n$=19}\\{\scriptsize $\mu$=298.45s [189.05, 323.00]}\\{\scriptsize $r$=2167 [126, 2758]}}
& \shortstack[l]{\textbf{TaskOutput}$^{\dagger}$\\{\scriptsize $n$=23}\\{\scriptsize $\mu$=396.54s [302.89, 603.49]}\\{\scriptsize $r$=1260 [124, 1331]}} \\

& Top 3
& \shortstack[l]{\textbf{Bash}$^{\dagger}$\\{\scriptsize $n$=1609}\\{\scriptsize $\mu$=2.06s [0.07, 0.30]}\\{\scriptsize $r$=410 [17, 123]}}
& \shortstack[l]{\textbf{Bash}$^{\dagger}$\\{\scriptsize $n$=3551}\\{\scriptsize $\mu$=1.94s [0.12, 0.30]}\\{\scriptsize $r$=212 [17, 117]}}
& \shortstack[l]{\textbf{ScheduleWakeup}\\{\scriptsize $n$=1}\\{\scriptsize $\mu$=20.67s [20.67, 20.67]}\\{\scriptsize $r$=41 [41, 41]}}
& \shortstack[l]{\textbf{Bash}$^{\dagger}$\\{\scriptsize $n$=4631}\\{\scriptsize $\mu$=4.87s [0.21, 1.05]}\\{\scriptsize $r$=398 [19, 372]}} \\

& Token-only extras
& {\scriptsize --}
& {\scriptsize --}
& {\scriptsize Grep ($r$=1379), Read ($r$=620)}
& {\scriptsize Read ($r$=675), Grep ($r$=258)} \\

\midrule

\multirow{4}{*}{\textbf{SWE-bench Pro}}
& Top 1
& \shortstack[l]{\textbf{Agent}$^{\dagger}$\\{\scriptsize $n$=3}\\{\scriptsize $\mu$=212.29s [35.20, 310.73]}\\{\scriptsize $r$=244 [118, 328]}}
& \shortstack[l]{\textbf{Agent}$^{\dagger}$\\{\scriptsize $n$=8}\\{\scriptsize $\mu$=67.35s [34.89, 60.47]}\\{\scriptsize $r$=623 [573, 704]}}
& \shortstack[l]{\textbf{TaskOutput}$^{\dagger}$\\{\scriptsize $n$=39}\\{\scriptsize $\mu$=90.65s [41.90, 120.98]}\\{\scriptsize $r$=459 [56, 472]}}
& \shortstack[l]{\textbf{TaskOutput}$^{\dagger}$\\{\scriptsize $n$=49}\\{\scriptsize $\mu$=75.57s [30.68, 87.94]}\\{\scriptsize $r$=345 [57, 278]}} \\

& Top 2
& \shortstack[l]{\textbf{TaskOutput}\\{\scriptsize $n$=22}\\{\scriptsize $\mu$=76.10s [40.77, 63.86]}\\{\scriptsize $r$=198 [65, 68]}}
& \shortstack[l]{\textbf{TaskOutput}$^{\dagger}$\\{\scriptsize $n$=4}\\{\scriptsize $\mu$=20.02s [6.37, 33.51]}\\{\scriptsize $r$=3872 [1011, 4264]}}
& \shortstack[l]{\textbf{Agent}$^{\dagger}$\\{\scriptsize $n$=75}\\{\scriptsize $\mu$=59.15s [42.43, 72.70]}\\{\scriptsize $r$=1662 [1259, 1881]}}
& \shortstack[l]{\textbf{Agent}$^{\dagger}$\\{\scriptsize $n$=89}\\{\scriptsize $\mu$=58.47s [40.35, 71.26]}\\{\scriptsize $r$=1776 [1287, 2186]}} \\

& Top 3
& \shortstack[l]{\textbf{Bash}$^{\dagger}$\\{\scriptsize $n$=691}\\{\scriptsize $\mu$=15.80s [0.09, 4.30]}\\{\scriptsize $r$=358 [26, 256]}}
& \shortstack[l]{\textbf{Bash}\\{\scriptsize $n$=646}\\{\scriptsize $\mu$=10.68s [0.25, 2.49]}\\{\scriptsize $r$=243 [43, 309]}}
& \shortstack[l]{\textbf{Bash}$^{\dagger}$\\{\scriptsize $n$=1473}\\{\scriptsize $\mu$=14.63s [0.08, 8.06]}\\{\scriptsize $r$=458 [58, 528]}}
& \shortstack[l]{\textbf{TaskStop}\\{\scriptsize $n$=17}\\{\scriptsize $\mu$=8.88s [0.04, 2.00]}\\{\scriptsize $r$=120 [84, 95]}} \\

& Token-only extras
& {\scriptsize Grep ($r$=295), TaskStop ($r$=269)}
& {\scriptsize EnterPlanMode ($r$=1157)}
& {\scriptsize Read ($r$=2519), Grep ($r$=342)}
& {\scriptsize Read ($r$=2232), Grep ($r$=246)} \\

\bottomrule
\end{tabularx}
\end{table*}

\noindent
\textbf{\underline{Top-3 Most Expensive Tool Calls.}}
Table~\ref{tab:rq9_rq11_combined_top_tools} summarizes the tools with the largest mean execution time for three representative workloads, while also marking tools that return large observations. Here, Agent refers to the delegation tool: a main-agent call that spawns a subagent to work on a bounded subproblem, rather than a direct shell, read, or edit action. Thus, when Agent appears as a top time-consuming tool, it indicates that a substantial portion of runtime is spent in delegated work. This is most visible in GAIA, where Agent takes 321.6s for Gemma-T, 916.0s for Qwen-T, and 650.7s for Qwen-I, suggesting that subagent-based reconnaissance, retrieval, or specialist analysis can dominate execution time.

The table also shows that expensive tools are often token-heavy. Many entries marked with $\dagger$ appear both among the slowest tools and among the largest result producers, meaning they affect serving twice: they stall agent progress and append large observations to the next-turn context. For example, on SWE-bench Pro, Agent returns 1,662--1,776 tokens for Qwen while taking roughly 58--59s per call, and TaskOutput returns 3,872 tokens for Gemma-I. Similarly, on ADE, Qwen’s Agent calls return over 1.6K tokens while taking more than 56s per call. This connects directly to Figure~\ref{fig:context_cache_breakdown}: even when tool execution occurs outside the LLM, its returned observations become newly appended input tokens, increasing context growth, KV-cache footprint, and future model-side cost.

\begin{table*}[t]
\centering
\footnotesize
\setlength{\tabcolsep}{3.2pt}
\renewcommand{\arraystretch}{1.08}
\caption{Top failing tool types with each cell reporting the tool type, call count $n$, and failure rate.}
\label{tab:failing_tools}
\begin{tabularx}{\textwidth}{llXXXX}
\toprule
\textbf{Benchmark} & \textbf{Rank} & \textbf{Gemma-T} & \textbf{Gemma-I} & \textbf{Qwen-T} & \textbf{Qwen-I} \\
\midrule
\multirow{3}{*}{\textbf{ADE}} & Top 1 & \shortstack[l]{\textbf{Bash}\\{\scriptsize $n$=208, fail=28.4\%}} & \shortstack[l]{\textbf{Edit}\\{\scriptsize $n$=2757, fail=95.4\%}} & \shortstack[l]{\textbf{Bash}\\{\scriptsize $n$=611, fail=18.5\%}} & \shortstack[l]{\textbf{Bash}\\{\scriptsize $n$=747, fail=16.6\%}} \\
 & Top 2 & \shortstack[l]{\textbf{Edit}\\{\scriptsize $n$=179, fail=11.7\%}} & \shortstack[l]{\textbf{Bash}\\{\scriptsize $n$=1462, fail=5.7\%}} & \shortstack[l]{\textbf{Edit}\\{\scriptsize $n$=189, fail=16.4\%}} & \shortstack[l]{\textbf{Edit}\\{\scriptsize $n$=186, fail=11.3\%}} \\
 & Top 3 & \shortstack[l]{\textbf{Read}\\{\scriptsize $n$=267, fail=1.5\%}} & \shortstack[l]{\textbf{Glob}\\{\scriptsize $n$=68, fail=1.5\%}} & \shortstack[l]{\textbf{Read}\\{\scriptsize $n$=836, fail=1.8\%}} & \shortstack[l]{\textbf{Grep}\\{\scriptsize $n$=109, fail=1.8\%}} \\
\midrule
\multirow{3}{*}{\textbf{GAIA}} & Top 1 & \shortstack[l]{\textbf{Read}\\{\scriptsize $n$=90, fail=48.9\%}} & \shortstack[l]{\textbf{Read}\\{\scriptsize $n$=28, fail=32.1\%}} & \shortstack[l]{\textbf{AskUserQuestion}\\{\scriptsize $n$=1, fail=100.0\%}} & \shortstack[l]{\textbf{Bash}\\{\scriptsize $n$=4631, fail=6.5\%}} \\
 & Top 2 & \shortstack[l]{\textbf{Bash}\\{\scriptsize $n$=1609, fail=7.1\%}} & \shortstack[l]{\textbf{WebFetch}\\{\scriptsize $n$=123, fail=7.3\%}} & \shortstack[l]{\textbf{SendMessage}\\{\scriptsize $n$=1, fail=100.0\%}} & \shortstack[l]{\textbf{WebFetch}\\{\scriptsize $n$=3761, fail=4.2\%}} \\
 & Top 3 & \shortstack[l]{\textbf{WebFetch}\\{\scriptsize $n$=1810, fail=3.0\%}} & \shortstack[l]{\textbf{Bash}\\{\scriptsize $n$=3551, fail=1.7\%}} & \shortstack[l]{\textbf{Read}\\{\scriptsize $n$=226, fail=14.6\%}} & \shortstack[l]{\textbf{Write}\\{\scriptsize $n$=85, fail=3.5\%}} \\
\midrule
\multirow{3}{*}{\textbf{SWE-bench Pro}} & Top 1 & \shortstack[l]{\textbf{TaskStop}\\{\scriptsize $n$=1, fail=100.0\%}} & \shortstack[l]{\textbf{ExitPlanMode}\\{\scriptsize $n$=1, fail=100.0\%}} & \shortstack[l]{\textbf{Bash}\\{\scriptsize $n$=1473, fail=7.1\%}} & \shortstack[l]{\textbf{ExitPlanMode}\\{\scriptsize $n$=2, fail=100.0\%}} \\
 & Top 2 & \shortstack[l]{\textbf{Edit}\\{\scriptsize $n$=984, fail=37.2\%}} & \shortstack[l]{\textbf{Edit}\\{\scriptsize $n$=955, fail=77.8\%}} & \shortstack[l]{\textbf{Edit}\\{\scriptsize $n$=661, fail=1.5\%}} & \shortstack[l]{\textbf{UpdateTodo}\\{\scriptsize $n$=1, fail=100.0\%}} \\
 & Top 3 & \shortstack[l]{\textbf{Bash}\\{\scriptsize $n$=691, fail=30.7\%}} & \shortstack[l]{\textbf{Bash}\\{\scriptsize $n$=646, fail=39.8\%}} & \shortstack[l]{\textbf{Write}\\{\scriptsize $n$=94, fail=1.1\%}} & \shortstack[l]{\textbf{Bash}\\{\scriptsize $n$=2820, fail=7.9\%}} \\
\bottomrule
\end{tabularx}
\end{table*}
\noindent
\textbf{\underline{Top-3 Failing Calls.}}
Table~\ref{tab:failing_tools} reports tool types with lowest success rates for the three representative workloads. The main takeaway is that \textit{failures concentrate in tools that modify state or execute code} (e.g., Bash and Edit). On ADE, Bash fails 28.4\% for Gemma-T and 16.6--18.5\% for Qwen, while Edit fails 95.4\% for Gemma-I. SWE-bench Pro shows a similar pattern for Gemma, where Edit fails 37.2--77.8\% and Bash fails 30.7--39.8\%. These failures arise from actions that require precise interaction with the environment, such as patching files, running commands, or depending on repository state, dependencies, and test behavior.

GAIA exposes a different failure mode. Read is the top failing tool for Gemma, failing 48.9\% for Gemma-T and 32.1\% for Gemma-I, while web-related tools such as WebFetch fail at lower but non-trivial rates. Some 100.0\% entries, such as AskUserQuestion, SendMessage, TaskStop, ExitPlanMode, and UpdateTodo, have only one or two calls and should therefore be interpreted cautiously. Compared with Table~\ref{tab:rq9_rq11_combined_top_tools}, these results show that expensive tools are not always the most failure-prone: Agent is often latency-heavy, whereas Bash, Edit, and Read are more prominent sources of retries and error observations. This helps explain the long tails in Figures~\ref{fig:turn_count_distribution} and~\ref{fig:context_usage_violin}: failed tool calls can trigger additional diagnosis and retries, increasing both repeated model invocations and accumulated context.

\begin{figure*}[t]
    \centering
    \includegraphics[width=\linewidth]{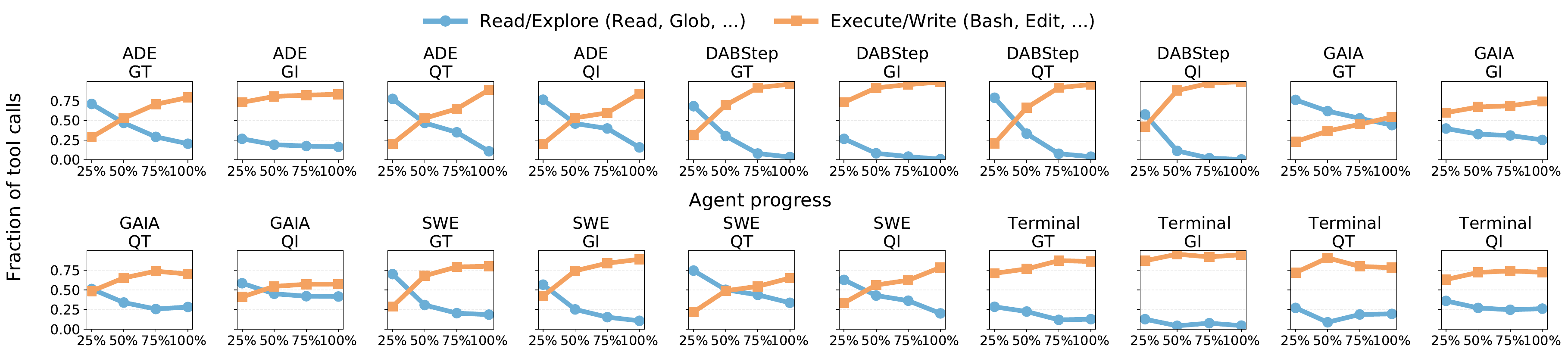}
    \caption{Progression of high-level tool intent over agent execution. Most agents shift from read/explore-heavy behavior early in execution to execute/write-heavy behavior later.}
\label{fig:tool_intention_progress_lines_combined}
\end{figure*}
\noindent
\textbf{\underline{Tool Intention Shift Over Time.}}
Tool calls are made with intentions. We classfiy Read, Glob, Grep, WebFetch, WebSearch as tool calls with \textit{Read/Explore} intention and Bash, Edit, Write, TodoWrite as the ones with \textit{Execute/Write} intention.
Figure~\ref{fig:tool_intention_progress_lines_combined} tracks how the fraction of tool calls devoted to these two intentions changes over the four quartiles of agent progress. A clear pattern emerges: \textit{agents typically begin in an inspection-heavy mode and then transition toward action-heavy behavior.} This transition is strongest on DABStep, SWE-bench Pro, and Terminal Bench. For example, in several DABStep and SWE-bench Pro configurations, Read/Explore starts near 0.6--0.8 of tool calls in the first quartile of turns and falls close to zero by the final quartile, while Execute/Write rises from roughly 0.2--0.4 to 0.8--1.0. Terminal Bench shows a similarly strong action-heavy late phase, with some configurations such as Gemma Instant remaining execute/write dominated almost throughout.

ADE follows the same overall trend, though with more model variation. Gemma Thinking and both Qwen variants show a visible crossover around the midpoint of execution, where Read/Explore declines and Execute/Write becomes dominant; by the final quartile, execute/write often accounts for roughly 0.7--0.9 of tool calls. Gemma Instant on ADE is a notable exception in the other direction: it is already heavily execute/write oriented in the first quartile, suggesting that this configuration enters action mode earlier rather than spending many turns on broad exploration. While Terminal bench exhibits a less significant trend, it's because most of its tool calls  are bash, which contains commands like `ls' with exploration intention.
By inspecting the specific bash commands, we do observe a more explicit trend.
The broader insight is that \textit{many agent trajectories appear to follow a natural pipeline: first inspect the environment or repository, then increasingly focus on acting on it.}

GAIA stands out as the main exception. Its curves are flatter and more balanced, with Read/Explore remaining substantial even late in execution and Execute/Write increasing only modestly. In configurations such as Gemma Thinking and Qwen Instant, the two fractions remain relatively close throughout the trajectory, ending near an approximate 40--60 split rather than converging to a strongly action-dominated phase. This is consistent with earlier results showing that GAIA is more retrieval-centric and tool-diverse: unlike software-engineering tasks, it often requires sustained information gathering instead of a clean transition into patching or test execution.

\textit{\noindent\textbf{\underline{Summary of Insights.}}
Tool use is not a single systems category: the same ``tool-call'' abstraction can correspond to lightweight file inspection, expensive shell execution, web retrieval, or delegated subagent work. A surprising finding is that the tools that dominate latency are not always the tools that fail most often: `Agent' and `TaskOutput' can be latency- and token-heavy, while `Bash', `Edit', and `Read' are more failure-prone and can trigger retries. Agents also exhibit a temporal structure, shifting from exploration early to execution/write behavior later; failed agents often make this shift too, but continue testing, debugging, or exhausting retrieval without reaching a correct terminal state.}
\section{Limitations and Future Work} 
Our study focuses on a representative set of models and mainstream ReAct-style agent execution, leaving several directions for broader characterization in future work.

\noindent
\textbf{\underline{Larger Models.}}
While larger models generally have stronger capabilities, some of the trends observed in our study still hold. For example, failed runs continue to consume more tokens. We evaluate GPT-5.5 on Terminal-Bench 2.0 and SWE-bench Pro, as shown in Table~\ref{tab:gpt55_failure_tokens}. \begin{table}[t] \centering \caption{Token usage of successful and failed runs with GPT-5.5.} \label{tab:gpt55_failure_tokens} \footnotesize \setlength{\tabcolsep}{3pt} \renewcommand{\arraystretch}{1.08} \begin{tabularx}{\columnwidth}{@{}lXrr@{}} \toprule \textbf{Model@Harness} & \textbf{Benchmark} & \textbf{Success} & \textbf{Fail} \\ \midrule GPT-5.5@Terminus-2 & Terminal-Bench 2.0 & 15,832.36 & 20,192.76 (+27.5\%) \\ GPT-5.5@Codex & SWE-bench Pro & 35,672.22 & 42,863.56 (+20.2\%) \\ \bottomrule \end{tabularx} \end{table} 
These results suggest that some workload trends persist even with larger models. Since our infrastructure will be open-sourced, these differences can be studied more extensively in future work. 

\noindent
\textbf{\underline{Advanced Orchestration Techniques.}}
Our study focuses on mainstream ReAct-style execution, including bounded one-level subagent delegation in Claude Code. Advanced orchestration techniques, including speculative or overlapped tool execution, dynamic context compaction or summarization, and branching or tree-search workflows, are outside the scope of this study. These techniques may change tool latency, context evolution, and KV-cache behavior, and we leave their characterization to future work.
\section{Related Work} \label{section:related_work}

Prior work evaluates LLM agents through interactive benchmarks, progress metrics, framework efficiency, and infrastructure cost analysis~\cite{liu2023agentbench,ma2024agentboard,gioacchini2024agentquest,xu2026agentrace}.
However, their focus is largely at the task, framework, or aggregate execution level, rather than on the fine-grained coupling between LLM serving and external tool execution.
In contrast, we reconstruct end-to-end agent trajectories and correlate per-request serving traces with tool-level behavior to characterize how agentic workloads stress serving systems over time.
Some other prior work spans foundational agent workflows such as ReAct, Reflexion, and LATS~\cite{yao2023react,shinn2023reflexion,zhou2024lats}, infrastructure-cost characterization of dynamic reasoning~\cite{kim2025costdynamicreasoning}, and recent CPU/OS-centric studies of agentic execution~\cite{raj2025cpucentricagentic,zheng2026agentcgroup}. We focus on ReAct-style agents instantiated with Claude Code and correlate LLM-serving traces with tool-level behavior.

\section{Conclusion}
This paper characterized ReAct-style agent workloads by connecting agent trajectories with LLM serving traces and tool-level behavior.
The central finding was that \textit{agentic serving was not simply long-prompt serving}: executions were long-tailed in both turn count and context growth, while effective context caching made the measured workload decode-dominated by reusing most prior context across turns.
This \textit{decode-bound behavior, however, represented an ideal cache regime}; in practical deployments, many concurrent agents and long multi-turn contexts can exceed GPU memory, forcing systems to choose between recomputation and offloading.
We also found that \textit{long-tail behavior could be amplified by failure-driven retry loops}, where repeated unsuccessful tool actions appended error observations and kept the agent interacting without meaningful progress. 
Finally, tool calls differed widely in latency, returned context, and failure behavior. 
Our characterization also exposed a clear temporal signal in agent behavior: \textit{agents consistently move from exploration-heavy tool use early in execution toward action-oriented editing, testing, and finalization later}.
Together, these results suggest that \textit{efficient agentic serving requires jointly managing model execution, persistent context cache, tool interaction, and failure recovery rather than optimizing any one component in isolation}.

\section*{Acknowledgment}
We used AI tools for code development, figure plotting, and text editing.
This work was supported in part by the Semiconductor Research Corporation (SRC) under the GRC AIHW grant, the AMD AI \& HPC Cluster Program, and Advanced Micro Devices, Inc. through the Funding Academic Research program.

\bibliographystyle{IEEEtran}
\bibliography{00_main}

\appendix
\section{Artifact Appendix}

\subsection{Abstract}

This artifact accompanies our paper ``Agentic AI Workload Characterization,'' which characterizes ReAct-style agents from both the LLM-serving and tool-execution perspectives.
Our study covers two open-weight models, Gemma4-31B and Qwen3.6-27B, each in Thinking and Instant modes, across
five agentic benchmarks.

The artifact contains both our trace collection/post-processing infrastructure and our collected data to support our paper.
It provides detailed instructions for environment setup, data preparation, experiment execution, trace post-processing, and figure generation.
We also provide our collected data for quick reproduction of the figures in our paper.

In particular, we focus on reproducing:
\begin{itemize}
\item A representative experiment using Qwen3.6-27B Thinking on ADE-Bench, including agent execution tracing and trace post-processing.
\item Figures~2--11 in the main paper using our collected and post-processed data for all 20 benchmark/model configurations.
\end{itemize}
The first part validates our complete tracing and analysis pipeline.
The second allows reviewers to confirm the paper's main results without re-collecting the data over several days on GPUs.

In addition to the above reproduction workflow for quick evaluation of the artifact, the artifact also provides complete instructions and scripts for collecting results for all 20 benchmark/model configurations.

\subsection{Artifact check-list (meta-information)}

{\small
\begin{itemize}
  \setlength{\itemsep}{0pt}
  \setlength{\parskip}{0pt}
  \setlength{\parsep}{0pt}
  \item {\bf Algorithm:} Trace correlation; Trace post-processing; workload analysis.
  \item {\bf Program:} Python~3.13, Bash, Docker, Harbor, vLLM, and Jaeger (dependencies are managed by \texttt{uv}).
  \item {\bf Compilation:} Not required.
  \item {\bf Models:} Qwen3.6-27B and Gemma4-31B; Thinking and Instant modes.
  \item {\bf Datasets:} ADE-Bench, DABStep, GAIA, SWE-bench Pro, and Terminal-Bench 2.0. 
  Our raw and post-processed traces for all experiments in the paper.
  \item {\bf Run-time environment:} Linux; \texttt{uv}; Docker; NVIDIA Container Toolkit.
  \item {\bf Hardware:} CPU-only for reproducing the figures in the paper; two H100 NVL 80-GiB GPUs for the trace collection.
  \item {\bf Run-time state:} Model weights and benchmark containers for trace collection.
  \item {\bf Execution:} Example trace collection/post-processing pipeline; figure reproduction.
  \item {\bf Metrics:} Turns; tokens; context reuse; LLM/tool and prefill/decode time; tool behavior.
  \item {\bf Output:} agent execution traces; figures in the paper.
  \item {\bf Experiments:} For quick evaluation: collect and post-process traces from Qwen3.6-27B Thinking running ADE-Bench; reproduce Figures 2–11.
  For full experiment: collect traces for all 20 configurations covered in the paper.
  \item {\bf Disk space required:} Approximately 200~GB.
  \item {\bf Time to prepare workflow:} Approximately 5--15 minutes after download.
  \item {\bf Time to complete the experiments:} For quick evaluation: 30--60 minutes for the quick evaluation on Qwen 3.6/ADE-bench; a few minutes to reproduce the figures from our collected data. For full reproduction: 50--60 hours on two H100 GPUs for all 20 configurations.
  \item {\bf Publicly available:} Yes.
  \item {\bf Code license:} CC BY 4.0.
  \item {\bf Data license:} CC BY 4.0.
  \item {\bf Workflow automation:} Bash and Python scripts; \texttt{uv}; Docker.
  \item {\bf DOI:} Code: \url{https://doi.org/10.6084/m9.figshare.33188466.v1};
  data: \url{https://doi.org/10.6084/m9.figshare.33188042.v1}.
\end{itemize}
}

\subsection{Description}

\subsubsection{How to access}

The source code of the artifact can be accessed at
\url{https://figshare.com/articles/conference_contribution/IISWC_2026_AE_Source_Code/33188466}.

We also provide our collected raw and post-processed data at \url{https://figshare.com/articles/dataset/IISWC_2026_AE_Data/33188042}.

The reviewer entry point is \path{README.ae.md} in the codebase, which includes an example agent trace collection/post-processing experiment and instructions for reproducing the figures in the paper.

To reproduce all data for all 20 configurations, readers should follow the instructions in the main \path{README.md}.

\subsubsection{Hardware dependencies}

\begin{itemize}
\item We conducted our experiments on a server with two NVIDIA H100 NVL GPUs (80~GiB each), connected via NVLink.
\item Agent trace collection uses two 80-GiB NVIDIA H100 NVL GPUs with tensor parallelism of 2.
Figure reproduction using our collected data requires no GPU.
\end{itemize}

\subsubsection{Software dependencies}

\begin{itemize}
  \item Linux and Python~3.13.
  \item \texttt{uv}, \texttt{curl}, \texttt{tar}, \texttt{md5sum}, and
  \texttt{jq}.  
  All Python dependencies are locked in \path{uv.lock}.
  \item Docker and NVIDIA Container Toolkit for fresh trace collection.
  \item vLLM 0.20.0 with the artifact's OTEL additions and Jaeger 1.57.
\end{itemize}

\subsubsection{Datasets}

The quick evaluation uses ADE-Bench as the agent task.

To reproduce the figures, we provide the data we collected for all 20 combinations of the five datasets and four model modes.

ADE-Bench/DABStep/GAIA/SWE-bench/Terminal-Bench-2.0 are used to reproduce the results of all experiments covered in this paper.

\subsubsection{Models}

Qwen/Qwen3.6-27B is used for the quick evaluation.

Qwen/Qwen3.6-27B and google/gemma-4-31B-it are used for the full reproduction of all results.
\subsection{Installation}

Download and extract \path{iiswc-2026-ae-main.zip}, enter the source root, and follow \path{README.ae.md}. 
Run \verb|uv sync --frozen| to install the exact
environment.  
Download \path{iiswc-2026-data-checkpoint.tar.gz} from the data DOI and follow \path{README.data.md} to verify and extract it.  

\subsection{Experiment Workflow}
\begin{enumerate}
  \item Download the source archive and data checkpoint.  Follow \path{README.ae.md} for environment setup and \path{README.data.md} to verify and extract the checkpoint.
  \item Reproduce Figures~2--11 by following \texttt{instructions/} \\
  \texttt{artifact-evaluation/iiswc-plots/README.md}.
  \item To validate agent trace collection and post-processing infra with Qwen3.6-27B Thinking on ADE-Bench, follow \texttt{instructions/artifact-evaluation/}\\
  \texttt{data-collection-curation/README.md}.
\end{enumerate}

\subsection{Evaluation and expected results}

For figure reproduction, we expect ten PDFs for Figures~2--11 that match the
figures in the paper, together with ten summary CSV files.  Figure~1 is a
conceptual diagram and is not generated by the artifact.

For fresh data collection and processing, we expect new Qwen3.6-27B
Thinking/ADE-Bench raw and processed data organized like the provided data.
The exact numbers may differ due to the stochastic nature of LLM inference.
\end{document}